# Topological water wave states in a one-dimensional structure

Zhaoju Yang[1], Fei Gao[1] and Baile Zhang[1,2] [(a)]

1 *Division of Physics and Applied Physics, School of Physical and Mathematical Sciences,*

*Nanyang Technological University, Singapore 637371, Singapore.*

2 *Centre for Disruptive Photonic Technologies,*

*Nanyang Technological University, Singapore 637371, Singapore.*

**Abstract -** Topological concepts have been introduced into electronic, photonic, and phononic systems, but have not been studied in surface-water-wave systems. Here we study a one-dimensional periodic resonant surface-water-wave system and demonstrate its topological transition. By selecting three different water depths, we can construct different types of water waves - shallow, intermediate and deep water waves. The periodic surface-water-wave system consists of an array of cylindrical water tanks connected with narrow water channels. As the width of connecting channel varies, the band diagram undergoes a topological transition which can be further characterized by Zak phase. This topological transition holds true for shallow, intermediate and deep water waves. However, the interface state at the boundary separating two topologically distinct arrays of water tanks can exhibit different bands for shallow, intermediate and deep water waves. Our work studies for the first time topological properties of water wave systems, and paves the way to potential management of water waves.

(a) E-mail: blzhang@ntu.edu.sg (corresponding author)

**Introduction.** – Water waves propagating through periodic structures can exhibit many interesting phenomena, such as band gaps in periodic lattice[1-3], super-lensing effect[4], refraction[5], cloaking[6], and others[7-8], most of which are inspired by rapid progress in photonic systems[9-11] in the last three decades. However these phenomena of water waves are still isolated from topological concepts of recent topological physics[12-15].

The study of topological physics was initiated in quantum Hall effect[12] and topological insulators[14,15]. Their special topological band structures give rise to the one-way propagation of quantum electronic waves as topologically protected edge states along boundaries of the systems. These edge states are robust against back-scattering from disorders because of the lack of backscatter channels. By constructing similar topological states in classical wave systems, the topological concepts have transformed the understanding of fundamental wave behaviors in photonics[16-25] and phononics[26-28].

Topological properties of two- or three- dimensional Bloch bands can be defined by topological invariants, which are calculated in terms of the well-known Berry phase[29] during the adiabatic motion of a particle across the Brillouin zone. For one-dimensional (1D) cases, the topological invariant of Bloch bands is the so-called Zak phase[13], which was proposed theoretically by J. Zak in 1989. This topological property can be found in the condensed matter Su-Schrieffer-Heeger (SSH) model[30] of polyacetylene and linearly conjugated diatomic polymers[31]. However, Zak phase had not been observed until 2013 when it was experimentally measured in condensed matter ultra-cold atoms[32]. Meanwhile, Zak phase has been introduced into photonic systems[33,34]. Recently, Zak phase was further introduced into 1D periodic acoustic systems[35].

The purpose of this report is to bring the concept of topology into the water wave systems by investigating an effective SSH model for water waves. By studying a 1D periodic array of cylindrical water tanks connected with narrow water channels, we demonstrate the existence of topological transition as the width of the connecting water channels varies. Through selecting three different water depths, we can construct different types of water waves - shallow, intermediate and deep water waves. By numerically calculating the band structures and Zak phases, we confirm the existence of topological transition for the shallow, intermediate and deep water waves. However, the interface state at the boundary separating two topologically distinct arrays of water tanks can exhibit different bands for shallow, intermediate and deep water waves.

**Model** – Inspired by the theoretical SSH[30] model as schematically shown in bottom of Fig. 1a, we design a 1D periodic system of cylindrical water tanks connected with narrow water channels for water waves. The unit cell is shown in the upper part of Fig. 1a. The lattice constant is $2a$. Each unit cell has two cylinder water tanks with the same radius $r = 0.4a$. The two connecting water channels between two water tanks have width $w + \Delta w$ for the one at two sides of the unit cell and $w - \Delta w$ for the middle one. The width is $w = 0.16a$ and $\Delta w$ is the modulation for dimerization. The tanks partly filled with water at depth $h$ constitute the 1D periodic water wave system. We consider linearized surface water waves[36], which obey the Helmholtz equation:

$$(\nabla^2 + k^2)\eta(r) = 0 \tag{1}$$

where $\eta(r)$ is the vertical displacement of water surface. The wave vector $k$ satisfies the dispersion relations for different types of water waves: $\omega^2 = ghk^2$ for shallow water waves, $\omega^2 = gk\tanh(kh)$ for intermediate water waves, and $\omega^2 = gk$ for deep water waves. The gravitational acceleration is $g = 9.8\ m/s^2$. The boundary condition at the walls of water tanks is $\vec{n} \cdot \nabla \eta = 0$. Because of the periodicity, we can apply the Bloch theorem and perform the numerical calculations based on finite element method (commercial software COMSOL) to obtain the band structures.

**Results.** - First, we set lattice constant $a$=0.2 m and the uniform water depth $h = 0.5\ a$. We first consider the dispersion of $\omega^2 = gk\tanh(kh)$ for intermediate water waves. Since there are two resonant tanks in one unit cell, hereafter we only consider the two-band model with two lowest water-wave eigen modes, whose vertical displacement patterns of water surface are nearly single valued in each cylindrical tank. Two examples of the patterns of the eigen modes in one unit cell are shown in Fig. 2a. By sweeping the parameter $\Delta w$ at momentum $k = \pi/2a$, we arrive at the result as shown in Fig. 1b. The band crossing of two bands indicates the band gap closing and reopening, which shows the topological transition point at $\Delta w = 0$ for the water wave system.

By choosing three values of $\Delta w = 0.5w, 0, -0.5w$, we numerically calculate the band structures and show the results in Fig. 2b-d, respectively. We can see that in three panels at the bottom of Fig. 2, the band gap closes and then reopens with decreasing $\Delta w$ from positive to negative values. This result shows the topological transition in agreement with the result in Fig. 1b.

To investigate the topological property of the band gaps, we can extend the concept of Zak phase[13], which was first developed from electronic system, to our 1D periodic water tanks. The Zak phase characterizes the topological property of the Bloch bands and can be viewed as the Berry's phase[29] picked up by a particle moving across the Brillouin zone. By expressing the normalized Bloch wave of water waves as $\psi_{n,k}(r)$, we can define the Zak phase for water waves through the Bloch function $u_{n,k}(r) = e^{-ikr}\psi_{n,k}(r)$ as

$$\varphi_{Zak} = i \int_{-\pi/2a}^{\pi/2a} \langle u_{n,k}|\partial_k|u_{n,k}\rangle \, dk, \qquad (2)$$

where $n$ is band index, $k$ is momentum, and $2a$ is lattice constant. Note that besides momentum and band index, the Bloch function is also a function of two-dimensional spatial positions.

After numerically calculating the Zak phase of the first band by using the cell-periodic Bloch function obtained from simulation, we find $\pi$ and $0$ as shown in blue and red numbers in panel b and d with $\Delta w = 0.5w, -0.5w$, respectively. Note that the Zak phase of each dimerization is a gauge dependent value, but the difference between the Zak phases of two dimerized configurations with $\Delta w > 0$ and $\Delta w < 0$, which is $\Delta \varphi_{Zak} = \varphi_{Zak1} - \varphi_{Zak2} = \pi$ in our water-wave model, is topologically defined[32]. The topological property of the band gap depends on the summation of the Zak phases of all the bands below the gap[14,15]. Therefore, band gaps in Fig. 2b, d are topologically distinct to each other.

In contrast, there are distortions of band structures for shallow water waves ($kh \ll 1$) and deep water waves ($kh \gg 1$) with respect to intermediate water waves. By setting $h = 0.05a$ and $h = 5a$, we arrive at the regions of shallow water waves and deep water waves. The calculated results of band structures with $\Delta w = 0.5w, 0, -0.5w$ as shown in Fig. 3a-c for shallow water waves and Fig. 3d-f for deep water waves. The blue, black and red dotted lines correspond to $\Delta w = 0.5w, 0, -0.5w$. We can see that in both cases there exists topological transition points between the first and second bands at $\Delta w = 0$. The Zak phases are shown in blue and red numbers. The results here are consistent with the intermediate water wave case, which verifies the existence of the topological transition for all three cases. Note that for the case of shallow water wave in contrast with intermediate case, there is a global frequency shift as shown in Fig. 3a-c and a little distortion which can be clearly seen in the interface-state calculations in Fig. 4a,c. For the deep water wave case, there is huge anomaly of eigenvalue calculation in which the eigenvalue approaches very large value at long wavelength limit ($k \rightarrow 0$). This is because in this limit of $k \rightarrow 0$, the system is no longer a deep-water-wave system, but changes to an intermediate-water-wave system and then a shallow-water-wave system with $kh \ll 1$. As a result, in Fig. 3d-f, the transition from shallow water waves to intermediate water waves and then deep water waves can be included all as the wave vector $k$ changes from 0 to $\pi/2a$.

Although the Zak phases characterize the geometric properties of the bulk bands, they can be used to determine the presence of the interface states located at the boundary between different structures[32,37]. The topologically nontrivial phases in Fig. 1,2 underlie the existence of protected interface states. In the previous water wave system as in Fig. 2, there will be interface states localized at the boundary between two kinds of semi-infinite topologically distinct water tank arrays. By choosing intermediate water wave model with water depth $h = 0.5a$, we calculated the band structure and demonstrate the vertical displacement of water surface for one interface state. In Fig. 4a, the interface is between two arrays of water tanks with width modulation $\Delta w = 0.5w$ and $\Delta w = -0.5w$, respectively. In view of the previous result in Fig. 1b, the two semi-infinite arrays of water tanks are topologically distinct to each other. We find that there is one interface state localized inside the band gap. Note that in calculation, we choose a finite structure and neglect the states located at two ends of the structure for simplicity. In Fig. 4b, the interface is between two arrays of water tanks with width modulation $\Delta w = 0.5w$ and $\Delta w = 0.3w$, respectively. As expected, there is no eigen state within the band gap. Furthermore, we compare the interface states under different dispersions as in Fig. 4c,d. The panels in the order of Fig. 4c,a,d show the differences of band structures of interface states from shallow water waves to intermediate water waves and then deep water waves. The interface states change from flat states inside a complete band gap (Fig. 4c) into dispersive state within a complete band gap (Fig. 4a), and then dispersive state within an incomplete bang gap (Fig. 4d). To visualize the interface states, we plot the three-dimensional pattern of vertical displacement of water surface in Fig. 4e from the results shown in Fig. 4a. The structure has total 20 unit cells (left: 10 unit cells with $\Delta w = -0.5w$ and right: 10 unit cells with $\Delta w = 0.5w$). Figure 4a-d all use the same structure. The red-blue pattern shows the positive and negative water surface displacement of the interface state. They strongly localize at the boundary and decay rapidly into the bulks. The green arrow indicates the location of the boundary. We should note that, if we excite the periodic water tanks in the band gap with the same parameters as in Fig. 4b, field attenuation from the excitation point can still be found because of the lack of bulk channels inside the band gap. However, this is not a localized interface state, and its position depends on the position of the excitation.

**Conclusions.** – In summary, we investigated the topological transition and extended the Zak phase[13] into a 1D water wave system. By calculating the band structures and Zak phases, we verify the topological transition for shallow, intermediate and deep water wave cases. We also demonstrated the interface states in one specific boundary separating two kinds of structure configurations. The parameters can be scaled up or down if needed. The measurement of Zak phase could use the reflection-phase method as shown in Ref. 35. The reflection phases of wave functions can be directly observed by taking photos. The vertical displacement of water surface for the interface state can also be observed directly. The results here bring the concept of topology into water wave systems in 1D system and may have potential applications in the future.

∗∗∗

**Acknowledgments**
This work was sponsored by Nanyang Technological University under Start-Up Grants, and Singapore Ministry of Education under Grant No. MOE2015-T2-1-070 and Grant No. MOE2011-T3-1-005.

**Author contributions**
B. L. Zhang supervised the project. Zhaoju Yang performed the modelling and calculations. All authors discussed and prepared the manuscript.

**Additional information**

Correspondence and requests for materials should be addressed to B. L. Zhang.

**Competing financial interests**

The authors declare no competing financial interests.

**References**

1. Chou, T. Liquid Surface Wave Band Structure Instabilities. Phys. Rev. Lett. 79, 4802 (1997).
2. Torres, M. et al. Visualization of Bloch waves and domain walls. Nature (London) 398, 114 (1999).
3. Hu, X. et al. Band structures and band gaps of liquid surface waves propagating through an infinite array of cylinders. Phys. Rev. E 68, 037 301 (2003).
4. Hu, X. et al. Superlensing effect in liquid surface waves. Phys. Rev. E 69, 030201(R) (2004).
5. Hu, X. and Chan, C. T. Refraction of water waves by periodic cylinder arrays. Phys. Rev. Lett. 95, 154501 (2005).
6. Farhat, M., Enoch, S., Guenneau, S. and Movchan, A. B. Broadband cylindrical acoustic cloak for linear surface waves in a fluid. Phys. Rev. Lett. 101, 134501 (2008).
7. Hu, X., Chan, C. T. Ho, K. and Zi, J. Negative effective gravity in water waves by periodic resonator arrays. Phys. Rev. Lett. 106, 174501 (2011).
8. Berraquero, C. P., Maurel, A., Petitjeans, P. and Pagneus, V. Experimental realization of a water-wave metamaterial shifter. Phys. Rev. E 88, 051002 (R) (2013).
9. Yablonovitch, E. Inhibited spontaneous emission in solid-state physics and electronics. Phys. Rev. Lett. 58, 2059 (1987).
10. John, S. Strong Localization of Photons in Certain Disordered Dielectric Superlattices. Phys. Rev. Lett. 58, 2486 (1987).
11. Chen, H., Chan, C. T. and Sheng, P. Transformation optics and metamaterials. Nature Mat. 9, 387-396 (2013).
12. Klitzing, K. V. The quantized Hall effect. Rev. Mod. Phys. 58, 519-531 (1986).
13. Zak, J. Berry's phase for energy bands in solids. Phys. Rev. Lett. 62, 2747-2750 (1989).
14. Hasan, M. Z. & Kane, C. L. Colloquium: Topological insulators. Rev. Mod. Phys. 82, 3045–3067 (2010).
15. Qi, X. L. & Zhang, S. C. Topological insulators and superconductors. Rev. Mod. Phys. 83, 1057-1110 (2011).
16. Haldane, F. & Raghu, S. Possible realization of directional optical waveguides in photonic crystals with broken time-reversal symmetry. Phys. Rev. Lett. 100, 013904 (2008).
17. Wang, Z., Chong, Y., Joannopoulos, J. & Soljacic, M. Reflection-free one-way edge modes in a gyromagnetic photonic crystal. Phys. Rev. Lett. 100, 13905 (2008).
18. Wang, Z., Chong, Y., Joannopoulos, J. D. & Soljacic, M. Observation of unidirectional backscattering-immune topological electromagnetic states. Nature 461, 772-775 (2009).
19. Hafezi, M., Demler, E. A., Lukin, M. D. & Taylor, J. M. Robust optical delay lines with topological protection. Nature Phys. 7, 907-912 (2011).
20. Poo, Y., Wu, R. X., Lin, Z. F., Yang, Y. & Chan, C. T. Experimental Realization of Self-Guiding Unidirectional Electromagnetic Edge States. Phys. Rev. Lett. 106, 093903 (2011).
21. Fang, K., Yu, Z. & Fan, S. Realizing effective magnetic field for photons by controlling the phase of dynamic modulation. Nature Photon. 6, 782-787 (2012).
22. Hafezi, M., Mittal, S., Fan, J., Migdall, A. & Taylor, J. M. Imaging topological edge states in silicon photonics. Nature Phot. 7, 1001-1005 (2013).
23. Rechtsman, M. C. et al. Photonic floquet topological insulators. Nature 496, 196-200 (2013).

24. Khanikaev, A. B. et al. Photonic topological insulators. Nature Mater. 12, 233-239 (2012).
25. Lu, L., Joannopoulos, J. D. & Soljacic, M. Topological photonics. Nature Phot. 8, 821-829 (2014).
26. Kane, C. L. & Lubensky, T. C. Topological boundary modes in isostatic lattices. Nature Phys. 10, 39-45 (2013).
27. Paulose, J., Chen, B. G. & Vitelli, V. Topological modes bound to dislocations in mechanical metamaterials. Nature Phys. 11, 153-156 (2014).
28. Yang, Z. et al. Topological acoustics. Phys. Rev. Lett. 114, 114301 (2015).
29. Berry, M. V. Quantal phase factors accompanying adiabatic changes. Proc. R. Soc. Lond. A 392, 45-57 (1984).
30. Su, W. P., Schrieffer, J. R. & Heeger, A. J. Solitons in polyacetylene. Phys. Rev. Lett. 42, 1698-1701 (1979).
31. Rice, M. J. & Mele, E. J. Elementary excitations of a linearly conjugated diatomic polymer. Phys. Rev. Lett. 49, 1455-1459 (1982).
32. Atala, M. et al. Direct measurement of the Zak phase in topological Bloch bands. Nature Phys. 9, 795-800 (2013).
33. Longhi, S. Zak phase of photons in optical waveguide lattices. Opt. Lett. 38, 3716-3719 (2013).
34. W. Tan, Y. Sun, H. Chen and S. Q. Shen. Photonic simulation of topological excitations in metamaterials. Scientific Reports 4, 3842 (2014).
35. Xiao, M. et al. Geometric phase and band inversion in periodic acoustic systems. Nature Phys. 11, 240-244 (2015).
36. Kundu, P. K. & Cohen, I. M. Fluid Mechanics. (Elsevier, USA, 2012).
37. Xiao, M., Zhang, ZQ. & Chan C. T. Surface impedance and bulk band geometric phases in one-dimensional systems. Phys. Rev. X 4, 021017 (2014).

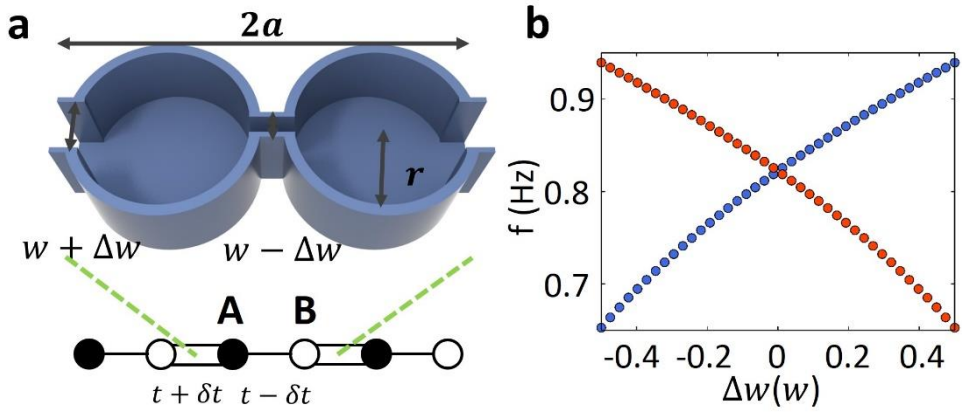

Figure 1. The periodic water tank structure and topological transition for water waves. **a,** the three dimensional view of one unit cell of water tank with lattice constant $2a$. **b,** the topological transition point at $\Delta w = 0$. Parameters: lattice constant $2a = 0.4$ m, radius of cylinder tank $r = 0.4a$, water depth $h = 0.5a$ (intermediate water waves), width of connecting rank $w = 0.16a$.

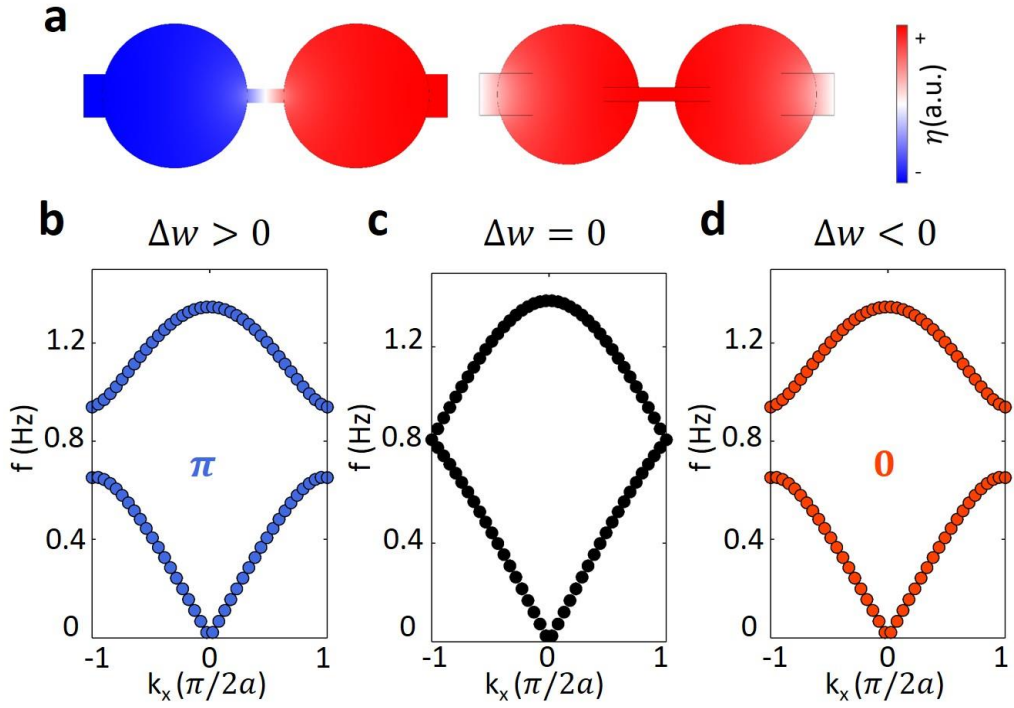

Figure 2. Eigen modes and band structures for intermediate water wave case. **a,** the two eigen modes with nearly single-valued vertical displacement of the water surface at $k_x = \pi/2a$ in panel b. The red-blue pattern shows the positive and negative water surface displacement. The water depth is $h = 0.5a$. Band structures of 1D periodic watery system with width parameters **b,** $\Delta w = 0.5w$ (blue dotted lines), **c,** $\Delta w = 0$ (black dotted lines), **d,** $\Delta w = -0.5w$ (red dotted lines). The first band gaps in panel b and d are topologically distinct to each other. The blue and red numbers in panel **b, d** represent the Zak phase. Panel **b** is the band diagram at the topological transition point. The other parameters are the same as in Figure 1.

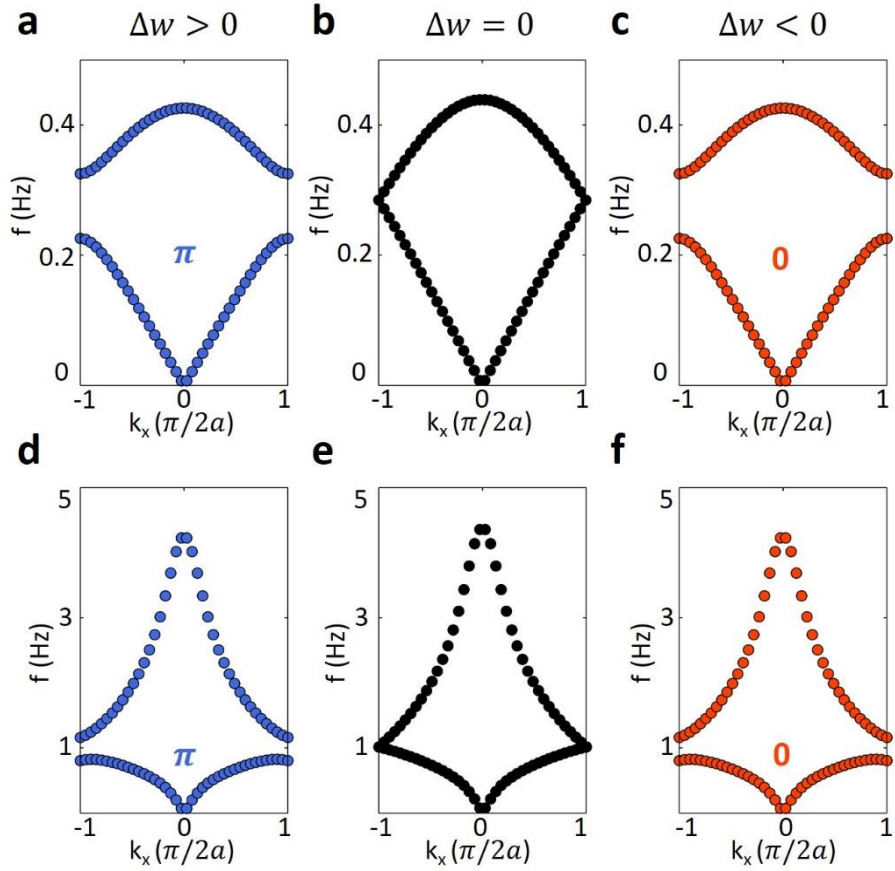

Figure 3. Band structures of shallow (water depth $h = 0.05a$) and deep (water depth $h = 5a$) water wave cases. **a-c,** three panels show the band diagrams for shallow water wave case with $\Delta w = 0.5w$ (blue dotted lines), $\Delta w = 0$ (black dotted lines) and $\Delta w = -0.5w$ (red dotted lines). **d-f,** three panels show the band structures for deep water wave case with $\Delta w = 0.5w$, $\Delta w = 0$ and $\Delta w = -0.5w$. In the limit of $k \to 0$, the system changes to an intermediate-water-wave system and then a shallow-water-wave system with $kh \ll 1$.

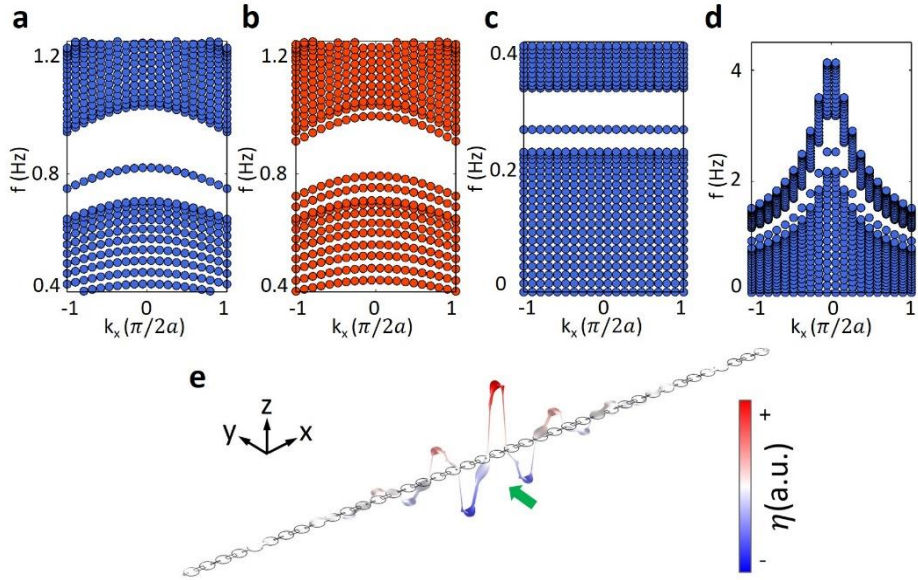

Figure 4 | Band diagrams of the water-wave systems with an interface. The band structure of two kinds of interface: **a,** between 10 unit cells with $\Delta w = 0.5w$ and 10 unit cells with $\Delta w = -0.5w$. **b,** between 10 unit cells with $\Delta w = 0.5w$ and 10 unit cells with $\Delta w = 0.3w$. There is an interface state localized inside the band gap in panel **a**, whereas no interface state in panel **b**. **c, d,** The nontrivial interface states for shallow water wave system $h=0.05a$ and deep water wave system $h=5a$. **e,** the vertical displacement of water surface of the interface state in panel **a**. The red-blue pattern shows the positive and negative surface displacement. Fig4. a,c,d all use the same structure. The other parameters are the same as in Figure 1.